# Tailoring the Electronic Configurations of YPc$_2$ on Cu(111): Decoupling Strategies for Molecular Spin Qubits Platforms

Soyoung Oh,[1,2,3] Franklin. H. Cho,[1,4] We-hyo Soe,[1,4] Jisoo Yu,[1,2] Hong Bui,[1,2] Lukas Spree,[1,4] Caroline Hommel,[1,4] Wonjun Jang,[1,4] Soo-hyon Phark,[1,4] Luciano Colazzo,[1,4] Christoph Wolf,[1,4,*] Fabio Donati[1,2,*]

[1]Center for Quantum Nanoscience, Institute for Basic Science (IBS), Seoul, 03760, Republic of Korea
[2]Department of Physics, Ewha Womans University, Seoul, 03760, Republic of Korea
[3]The Clarendon Laboratory, Department of Physics, University of Oxford, Oxford OX1 3PU, UK
[4]Ewha Womans University, Seoul, 03760, Republic of Korea
[*]Email: wolf.christoph@qns.science, donati.fabio@qns.science

## Abstract

Molecule-based spin architectures have been proposed as promising platforms for quantum computing. Among the potential spin qubit candidates, yttrium phthalocyanine double-decker (YPc$_2$) features a diamagnetic metal ion core that stabilizes the molecular structure, while its magnetic properties arise primarily from an unpaired electron (S=1/2) delocalized over the two phthalocyanine (Pc) ligands. Understanding its properties in the proximity of metal electrodes is crucial to assess its potential use in molecular spin qubits architectures. Here, we investigated the morphology and electronic structure of this molecule adsorbed on Cu(111) surface using scanning tunneling microscopy (STM). On Cu(111), YPc$_2$ adsorbs flat, with isolated molecules showing a preferred orientation along the ⟨111⟩ crystal axes. Moreover, we observed two different types of self-assembly molecular packing when growing molecular patches. For YPc$_2$ in direct contact with Cu(111), STM revealed a widely separated highest occupied and lowest unoccupied molecular orbitals (HOMO/LUMO), suggesting the quenching of the unpaired spin. Conversely, when YPc$_2$ is separated from the metal substrate by a few-layer thick diamagnetic zinc phthalocyanine (ZnPc) layer, we found the HOMO to split into singly occupied and singly unoccupied molecular orbitals (SOMO/SUMO). We observed that more than 2 layers of ZnPc are needed to avoid intermixing between the two molecules and spin quenching in YPc$_2$. Density functional theory (DFT) reveals the spin quenching is due to the hybridization between YPc$_2$ and Cu(111) states, confirming the importance of using suitable decoupling layers to preserve the unpaired molecular spin. Our results suggest the potential of YPc$_2$/ZnPc heterostructures as a stable and effective molecular spin qubit platform and validate the possibility of integrating this molecular spin qubit candidate in future quantum logic devices.

# 1. Introduction

Molecular spin qubits offer inherent discrete energy levels to realize scalable qubit platforms.[1, 2] This molecule-based architecture allows for qubit design through chemical approaches, enabling precise tailoring of the electronic structure, molecular spacing, and spin noise from the environment.[3, 4] Best molecular spin qubits exhibit long spin relaxation time ($T_1$) of tens of ms and spin coherence time ($T_2$) of hundreds of μs.[5-7] While the molecular spins in magnetically diluted bulk and solution states can be conveniently measured using ensemble-averaging techniques, difficulties arise in addressing single qubits individually.[1]

Conversely, molecular spins adsorbed on a surface provide a viable way to readout such molecules at the atomic scale using scanning probe techniques.[8, 9] Moreover, the capability of molecular spins to self-assemble on single crystal surfaces[10, 11] makes them stable and tunable building blocks for nanofabricating quantum devices.[12] Both molecule-surface and intermolecular interactions significantly impact the molecular spin states, hence the investigation of these properties is essential to achieve precise control of the molecular spin configurations. For example, molecular packing and adsorption site on the surface can regulate the charge transfer and electronic structure.[13, 14] Similarly, the electronic structure can also change depending on whether the molecules are in direct contact with a metal substrate or buffered by a diamagnetic layer.[15] The latter strategy can additionally serve to reduce the coupling between the molecular spin and the substrate{Cimatti, 2019 #694}, which is known to be a source of decoherence. For both the buffer layer and the molecular spins, a flat structure and a proper symmetry of the molecule are desirable properties for a regular assembly of the molecular layer.[16, 17]

Metal phthalocyanine double-decker ($MPc_2$) is one of the prevalent platforms of molecular spin qubits on a surface with the aforementioned properties.[18-21] The flat geometry of $MPc_2$ facilitates great structural stability, which contributes to the flat absorption and self-assembly on metal substrates. Furthermore, $MPc_2$ is robust during thermal sublimation in ultra-high vacuum (UHV).[8, 22] When a lanthanide metal (*e.g.*, Tb, Dy, Er) is used,[23] such molecules exhibit very long $T_1$ and $T_2$,[24, 25] as well as the single-molecule magnet behavior with enhanced magnetic stability.[26, 27] However, lanthanide phthalocyanine double-deckers consist of two exchange-coupled spin systems; one is the high-spin from 4f electrons localized on the lanthanide metal ion, and the other is the spin $S=1/2$ π-radical delocalized over the two phthalocyanine (Pc) ligands,[22] resulting in a complex magnetic structure with large spin multiplicity that can cause qubit leakage at high temperature.[6, 28] In addition, coherently controllable electron spin states of such phthalocyanine double-deckers are primarily limited to the radical spin transition due to the strong magnetic anisotropy of most lanthanides, making them less ideal as molecular qubit candidate.[18] A more suitable system can be created by replacing the lanthanides with diamagnetic ions. In particular, yttrium phthalocyanine double-decker ($YPc_2$) has a diamagnetic trivalent yttrium ion that stabilizes a single, quasi-isotropic unpaired electron in the Pc ligands,[29, 30] endowing a simple and clear understanding of the spin dynamics of a nearly-free electron. Due to its low spin-orbit coupling and weak hyperfine interaction with the nuclear spins in the molecule, $YPc_2$ exhibits a single and narrow electron spin resonance (ESR) spectrum, both in dense powder,[30] single crystals,[18] and few-nm thick molecular films.[31] Revealing its spin properties in the form of a self-assembled film on a surface requires the use surface-sensitive ensemble,[31-33] or scanning tunnelling microscope (STM) combined ESR setup.[31-33] However, it is crucial to first assess the feasibility of such characterizations by investigating the morphology and electronic properties of $YPc_2$ on single crystal surfaces, which can serve as a support for the molecular film in these types of setups.

Here, we investigate the morphology and electronic configuration of YPc$_2$ deposited directly on a Cu(111) and with a few-layer thick zinc phthalocyanine (ZnPc) decoupling layer inserted between Cu(111) and YPc$_2$. Specifically, we chose Cu(111) in view of characterizing these molecules using our recently developed surface-sensitive ensemble ESR setup,[31] whose functioning is presently optimized for this type of surface, while we used ZnPc as a diamagnetic spacer to tune the hybridization between YPc$_2$ and the Cu surface. In all cases, we observe flat adsorption of the YPc$_2$, with the Pc plane parallel to the surface. On Cu(111), the mismatch between the four-fold symmetry of the molecule and the global three-fold symmetry of the surface is reflected in the misalignment between the molecular lattice axis and the crystal lattice orientation. When only a single monolayer (ML) of ZnPc is grown on Cu(111), the subsequently deposited YPc$_2$ frequently embeds within the ZnPc layer. However, no intermixing is found when YPc$_2$ is deposited on 2 or more ML of ZnPc. Scanning tunneling spectroscopy (STS) reveals a large gap of more than 2.3 eV between the highest occupied and lowest unoccupied molecular orbitals (HOMO/LUMO) when YPc$_2$ is directly adsorbed on the Cu(111) and embedded into the ZnPc layer. Conversely, when adsorbed on top of another YPc$_2$ layer or on top 2 ML ZnPc, the HOMO splits into singly occupied and singly unoccupied molecular orbitals (SOMO/SUMO), while the gap is largely reduced. Combining these results with density functional theory (DFT) we find that, when YPc$_2$ is directly adsorbed on Cu(111), the hybridization between molecular orbitals and surface electrons quenches the unpaired electron, while a pristine electronic configuration with S=1/2 is attained when YPc$_2$ is decoupled from the metal using the ZnPc layer. Our results highlight the potential of YPc$_2$/ZnPc heterostructures as a robust molecular spin qubit architecture, contingent upon the ability to control its electronic configuration through a molecular decoupling layer.

## 2. Experimental and computational methods

Samples were prepared and measured using a home-built and a commercial STM (PanScan Freedom, RHK Technology). The former is equipped with an Auger electron spectrometer and was used to characterize the growth and composition of the molecules upon deposition on the Cu(111) surface. The latter was used to investigate the electronic structure of the surface-adsorbed molecules. In both machines, the preparation and STM chamber are connected, enabling the transfer of samples in UHV. This ensures that the sample surface is unchanged while transferring, allowing reproducible results. The Cu(111) surface is cleaned in the preparation chamber (base pressure ~ 1 x 10$^{-9}$ torr) by several cycles of Ar$^+$ sputtering (pressure 5 × 10$^{-6}$ torr, energy 1 keV) followed by thermal annealing around 673 K. The quality and (111) surface termination of the substrate were confirmed by STM measurements before molecule deposition. The YPc$_2$ molecules were synthesized in-house following a solvothermal route[34]. In the process, 2.685 g 1,2-Dicyanobenzene (8 eq), 0.844 g Yttrium acetate monohydrate (1 eq), and 1.588 g 1,8-Diazabicyclo [5.4.0] undec-7-ene (4 eq) were mixed in 30 mL Ethanol and heated in a Teflon lined autoclave at 190 °C for 12 hours. The mixture was allowed to cool down to room temperature over the course of 24 hours. The crude product was washed with n-hexane and diethylether on a glass filter frit. The resulting powder was mixed with 0.5666 g of periodic acid (H$_5$IO$_6$) and stirred overnight in 400 mL of a 1:1 mixture of chloroform and methanol to oxidize anionic YPc$_2$ species. Filtration and washing with 1:1 CHCl$_3$/MeOH over a glass filter frit were employed to remove insoluble byproducts like free base Pc. The solution was then reduced in volume under vacuum and dried with 30 g of activated alumina. Purification of the product was facilitated by column chromatography on deactivated alumina with 9:1 $CHCl_3$/MeOH and collecting the first green band eluting from the column. The identity of the target compound was confirmed by UV-Vis spectroscopy and mass spectrometry. The ZnPc molecules were purchased from Sigma-Aldrich (97 % purity). Both compounds were thermally sublimated on the sample kept at room temperature from a heated crucible. For the home-built STM, deposition of $YPc_2$ is performed using a home-built evaporator at 723–733 K, where the temperature of the crucible was measured using a K-type thermocouple wire positioned inside the crucible. For the RHK STM, we use a Kentax 3-cell thermal evaporator (TCE-BSC) at 633–653 K. Due to the different distances from the sample, the deposition temperatures in the two systems were adjusted to obtain a sublimation rate of ~ 0.01 ML/min. For the ZnPc, we set the sublimation temperature at 623 K to obtain a rate of 0.1 ML/min.

All STM and STS measurements were performed at 10 K using the RHK STM at a base pressure $< 3 \times 10^{-10}$ torr, except the data in **Figure 1** (a), which was acquired using the home-built STM operating at 120 K with a base pressure $< 4 \times 10^{-10}$ torr. Images were taken in a constant current mode, while differential conductance (dI/dV) spectra were obtained using a lock-in amplifier in constant height mode. The voltage bias is swept with added AC bias modulation of 20– ~ 50 mV. The dI/dV maps were measured with a modulation of 50 mV turned on in constant current mode. We scanned at a speed of 1–1 ~ 2 nm/s, allowing the tip to stay over the spatial region of a single pixel long enough to obtain the current signal with an integration time of 100– ~ 200 ms. As a result, we acquired the topographic information and electronic configuration of the molecules at the same time. All STM images are filtered by WSxM software.

Ab initio calculations were performed using plane-wave DFT with pseudopotentials as implemented in Quantum Espresso (V 7.3). The cutoff for the basis was set to 60 Ry (600 Ry for the charge density), dispersive forces were treated with the revised VV10 method and only the Gamma point was used in the integration of the Brillouin zone. Slab systems were built based on 3 ML of Cu(111), up to 4 ZnPc molecules and a single $YPc_2$ molecule, padded by ~2 nm of vacuum in z-direction. Molecules in vacuum were calculated in a $3 \times 3 \times 2$ nm vacuum box. All systems were relaxed before calculating the electronic ground state. To calculate the ionization potential (electron affinity) we use the Delta SCF method by adding or removing one electron whilst the molecules are frozen at the geometry of the neutral system.

## 3. Results and discussion

### 3.1 $YPc_2$ on Cu(111)

The top view of the molecular structure of $YPc_2$ is shown in **Figure 1** (a). The top Pc is represented by blue-colored nitrogen and black-colored carbon atoms, while the bottom Pc is indicated by red-colored nitrogen and pink-colored carbon atoms. Each Pc has four reflection planes, rotated by 45° relative to each other. We marked one representative symmetry plane per each Pc, $M_1$ for the top Pc and $M_2$ for the bottom Pc. The yttrium ion serves as a joint between the twisted Pc ligands. As a result, $YPc_2$ exhibits 4-fold $D_{4d}$ symmetry.

We first characterized the orientation of isolated YPc$_2$ on Cu(111) (**Figure 1** (b)) and how they self-assemble into molecular patches on Cu(111) (**Figure 1** (e)–(f)) with respect to the ⟨111⟩ surface lattice vectors using STM. We determined the latter using dislocation line defects generated by poking the STM tip into the surface, while we take the azimuthal rotational angle of M$_1$ with respect to the horizontal scan direction of the STM images. As shown in **Figure 1** (c), isolated YPc$_2$ molecules do not adsorb randomly on the surface but rather align along specific angles spaced at 30° intervals. These preferred M$_1$ angles correspond to the three ⟨111⟩ surface lattice vectors, see **Figure 1** (d). This result is in line with previous study of metal-free Pc on Au(111).

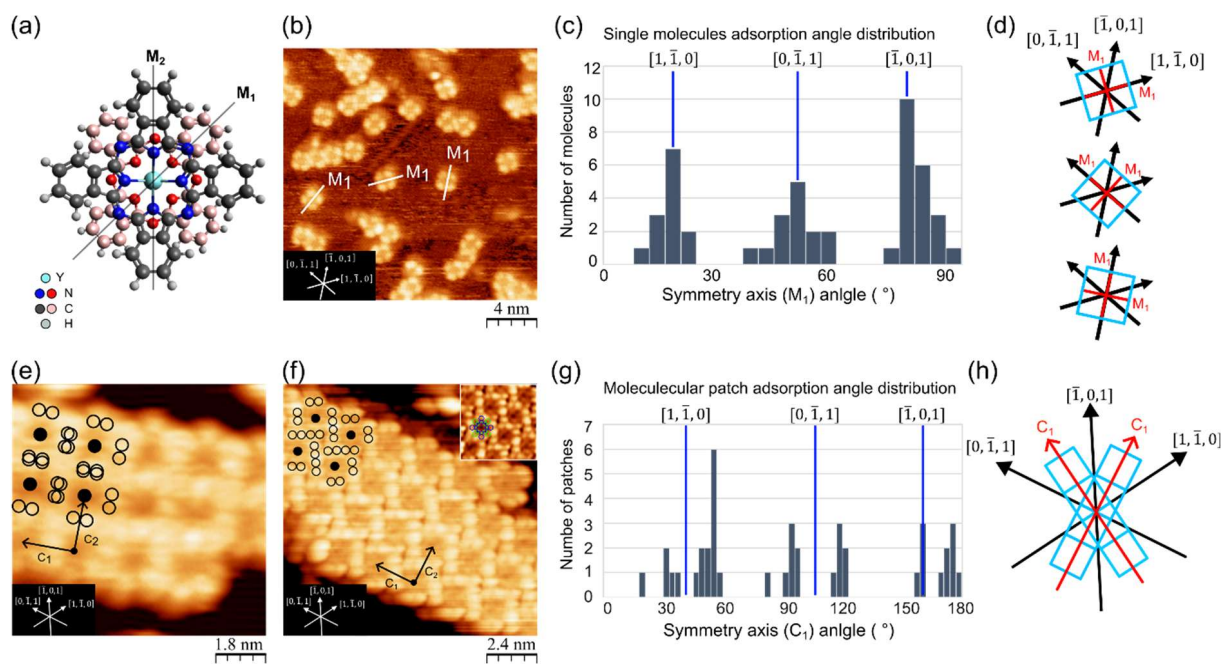

**Figure 1** Adsorption of YPc$_2$ on Cu(111). (a) YPc$_2$ molecular structure. Carbon and nitrogen atoms on the top (bottom) Pc are indicated with black (pink) and blue (red) circles, respectively. We mark one reflection plane passing through the corner nitrogen atoms of the top Pc as M$_1$, and similarly mark the same plane of the bottom Pc as M$_2$. (b) STM image (20 × 20 nm, V$_{DC}$ = 2.0 V, I$_{set}$ = 50 pA) of isolated YPc$_2$ molecules on Cu(111). White lines mark the reflection plane M$_1$. (c) Molecular orientation distribution obtained from 48 isolated molecules on Cu(111). The angle between M$_1$ and the horizontal direction of the STM image is indicated on the x-axis of the histogram with a bin size of 4°. Due to the 4-fold square symmetry of isolated YPc$_2$ molecule, the angle of M$_1$ for the adsorption is represented between 0° and 90°. Blue lines mark the surface lattice vectors of the (111) surface. (d) Schematics representing three patterns of YPc$_2$ molecules adsorbed along the ⟨111⟩ surface lattice vectors. (e) STM image (9 × 9 nm, V$_{DC}$ = -1.0 V, I$_{set}$ = 50 pA) of a YPc$_2$ molecular patch phase I on Cu(111). The patch unit vectors C$_1$ and C$_2$ are shown with black arrows. The position of the yttrium ion is marked with a filled black circle, while the 8 outer lobes are marked with hollow black circles. (f) STM image (12 × 12 nm, V$_{DC}$ = 1.0 V, I$_{set}$ = 20 pA) of a YPc$_2$ molecular patch phase II on Cu(111). The inset provides a magnified view of YPc$_2$ structure, with the chemical schematics of the top and bottom Pc rings indicated in blue and green, respectively. (g) Adsorption angle distribution obtained from 41 molecular patches on Cu(111). The angle between C$_1$ and the horizontal STM image direction is indicated with a bin size of 3°. Due to the absence of four-fold symmetry, the x-axis of the histogram ranges from 0° to 180°. (h) Alignment pattern of C$_1$ with respect to the surface lattice vectors.

Increasing the coverage of YPc$_2$ leads to formation of molecular patches with two types of self-assembled domain. We find one domain showing molecules in checkerboard patches without visible internal lobe structure (**Figure 1** (e)), while the other domain has molecules showing visible internal lobe structure (**Figure 1** (f)). In line with the previous work,[22] we label these two domains as phase I and II, respectively. The more abundant phase I is characterized by a dark molecular center with no discernible internal structure and molecular lobes of the top Pc overlapping with nearby ones. Conversely, for phase II, the molecular lobes of the top Pc avoid overlapping with each other, resulting in a side-to-side self-assembled arrangement. We determine the orientation of the molecular patches by comparing the molecular unit lattice vectors $C_1$ and $C_2$ with respect to the ⟨111⟩ surface lattice vectors, with focus on the axis $C_1$ that corresponds to the direction of maximal elongation of the patch. As shown in **Figure 1** (g), the distribution of the azimuthal rotational angle is peaked at specific angles deviating by approximately ±10–15° from each ⟨111⟩ surface lattice vector, which is schematically displayed in **Figure 1** (h).

To investigate the electronic properties of isolated YPc$_2$ on Cu(111), we perform point dI/dV spectroscopy and dI/dV maps. For the molecule shown in **Figure 2** (a), both dI/dV spectra at the molecular lobe and center sites show two clear peaks, one below and the other above the Fermi level indicating the HOMO and LUMO, respectively (**Figure 2** (b)). As marked by the vertical lines, the HOMO peak appears at the same energy (-0.7 eV) at both the lobe and center, while the LUMO peak at the molecule center (1.65 eV) is slightly shifted towards higher energy compared to that at the lobe (1.55 eV). As a consequence, the HOMO-LUMO gap at the center site (2.35 eV) is slightly larger than at the lobe (2.25 eV). The gap difference between the lobe and center suggests that the largest contribution to the conductance is due to the electron states at the Pc ligands, with a smaller contribution from the center yttrium ion.[35] The electronic structure of an isolated YPc$_2$ on Cu(111) is quite different from that of YPc$_2$ on Au(111), which shows 4 peaks and much narrower gap.[22] This result indicates that YPc$_2$ on Cu(111) has different electronic properties compared to YPc$_2$ on Au(111), which was shown to preserve the unpaired spin in the molecular ligands.[22] In addition, no Kondo resonance was detected at the Fermi level for YPc$_2$ on Cu(111) down to 10 K, which may either indicate the quenching of the unpaired spin or a Kondo screening occurring at lower temperatures.

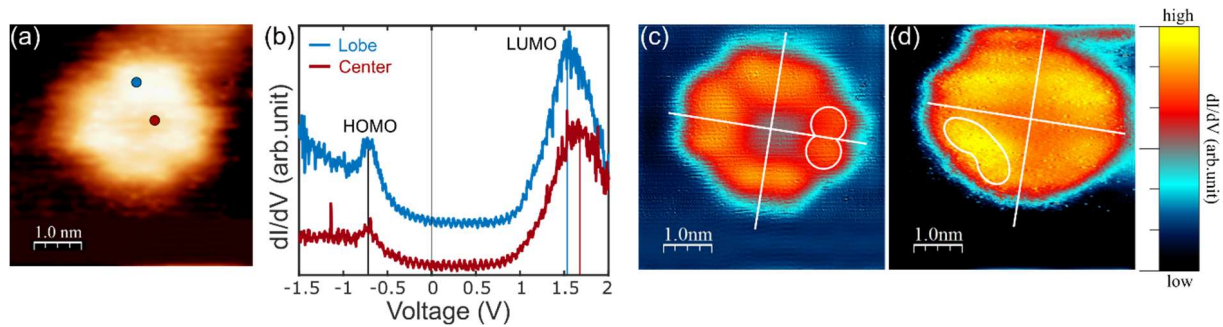

**Figure 2** Electronic properties of individual YPc$_2$ on Cu(111). (a) STM image (5 × 5nm, $V_{DC}$ = 1.0 V, $I_{set}$ = 50 pA) of isolated YPc$_2$ on Cu(111).5 nm0 V. (b) dI/dV spectra with the initial condition: $V_{DC}$ = 2.0 V, $I_{set}$ = 150 pA, voltage bias modulation ($V_{AC}$) = 50 mV. Spectra of the molecular lobe and center were acquired at the positions marked with blue and red dots in (a). Vertical solid lines mark the HOMO and LUMO peaks maxima in each spectrum. (c) dI/dV map measured at the HOMO energy, $V_{DC}$ = -0.7 V, $I_{set}$ = 50 pA, $V_{AC}$ = 50 mV. (d) dI/dV map measured at the LUMO energy, $V_{DC}$ = 1.4 V, $I_{set}$ = 50 pA, $V_{AC}$ = 50 mV. Each dI/dV map's intensity is rescaled independently for maximum contrast.

The differential conductance map of isolated YPc$_2$ on Cu(111) at the HOMO and LUMO energy are shown in **Figure 2** (c) and (d). Both maps of HOMO and LUMO are 4-fold symmetric, with their more pronounced minima rotated by 45°. The dI/dV maps disclose a high density of states at the Pc ring (lobe) site, confirming the smaller contribution of the center ion. Similar patterns and a large size gap over 2 eV in electronic structures have been observed in several vanadyl-based metal organic complexes with no spin radical on the ligand, such as VOPc on TiOPc/Ag(100)[16] and Au(111),[36] as well as calculated for VOTTDPz on Au(111).[37]

In order to investigate the role of inter-molecule interaction on their electronic configuration, we performed STS of YPc$_2$ patches on Cu(111) over the 4 sites marked in **Figure 3** (a) named as side lobe, center, surrounded, and attaching lobe. The dI/dV spectra acquired at these sites are shown in **Figure 3** (b). Similar to the isolated molecules, all spectra acquired over the molecular patch show the typical two-peaked HOMO-LUMO structure, however, both peaks appear at slightly lower energy compared to the isolated molecule. The HOMO peak maxima are consistent for all sites (-0.8 eV) except for the attaching lobe site (-0.9 eV). In contrast, the LUMO peak exhibits a site-dependent energy spread, with the lowest energy at the side lobe (1.35 eV) and the highest energy at the attaching point (1.45 eV). The reason for the HOMO-LUMO peak redshift upon formation of self-assembled molecular patches can be attributed to the formation of additional bonding states through inter-molecule orbital hybridization, which can broaden the density of states and alter the electronic structure. In addition, the local electronic structure depends on the interplay between intermolecular and substrate-molecule hybridization, which results in a site-dependent HOMO and LUMO peak. Nevertheless, in all cases the HOMO-LUMO gap remains larger than 2.0 eV, which is close to that of the isolated YPc$_2$ on Cu(111). Similar as in the isolated YPc$_2$, we observe no Kondo peak from the molecular patch.

To further assess the effect of the inter-molecule hybridization, we compare the dI/dV map of both isolated and self-assembled molecular patch at the HOMO and LUMO energies (**Figure 3** (c) and (d)). While the intensities and spatial distribution of the HOMO of isolated molecule and molecular patch are similar, the dI/dV map at the LUMO energy shows pronounced differences. As shown in **Figure 3** (d), isolated molecules show a higher conductance, but the conductance decreases when the structure progresses from individual molecules to linear chains, and to more extended patches. The bonding between ligands facilitates the delocalization of π-electrons, resulting in a red-shift and broadening of the unoccupied state.[35, 38]

As discussed above, the different electronic structure between YPc$_2$ on Cu(111) and Au(111), together with the absence of Kondo feature on Cu(111) point towards a quenching of the unpaired spin, possibly due to the proximity of the more reactive Cu(111) surface. To validate this conclusion, we performed dI/dV spectroscopy on a YPc$_2$ molecule adsorbed on a patch of YPc$_2$ on Cu(111), see **Figure 4** (a). Unlike the YPc$_2$ directly adsorbed on Cu(111), the dI/dV spectrum of the YPc$_2$ on top of a YPc$_2$ patch (**Figure 4** (b)) shows four distinct peaks, whose maxima are marked with vertical lines at: (1) -1.4 eV, (2) -0.75 eV, (3) 0.50 eV, (4) and 1.70 eV. In addition, the electronic gap of YPc$_2$ on YPc$_2$/Cu(111), *i.e.*, the difference between (2) and (3), is 1.25 eV, which is remarkably smaller than YPc$_2$ on Cu(111). The number of peaks and narrower gap are reminiscent of the electronic configuration on Au(111), for which the spin is not quenched. Different from YPc$_2$/Au(111), however, we do not observe any Kondo features on YPc$_2$ on YPc$_2$/Cu(111). For the latter, this might be due to a Kondo temperature that is much lower than the experimental temperature, possibly due to the presence of the 1$^{st}$ YPc$_2$ layer that screens the interaction with the Cu(111) conduction electrons.

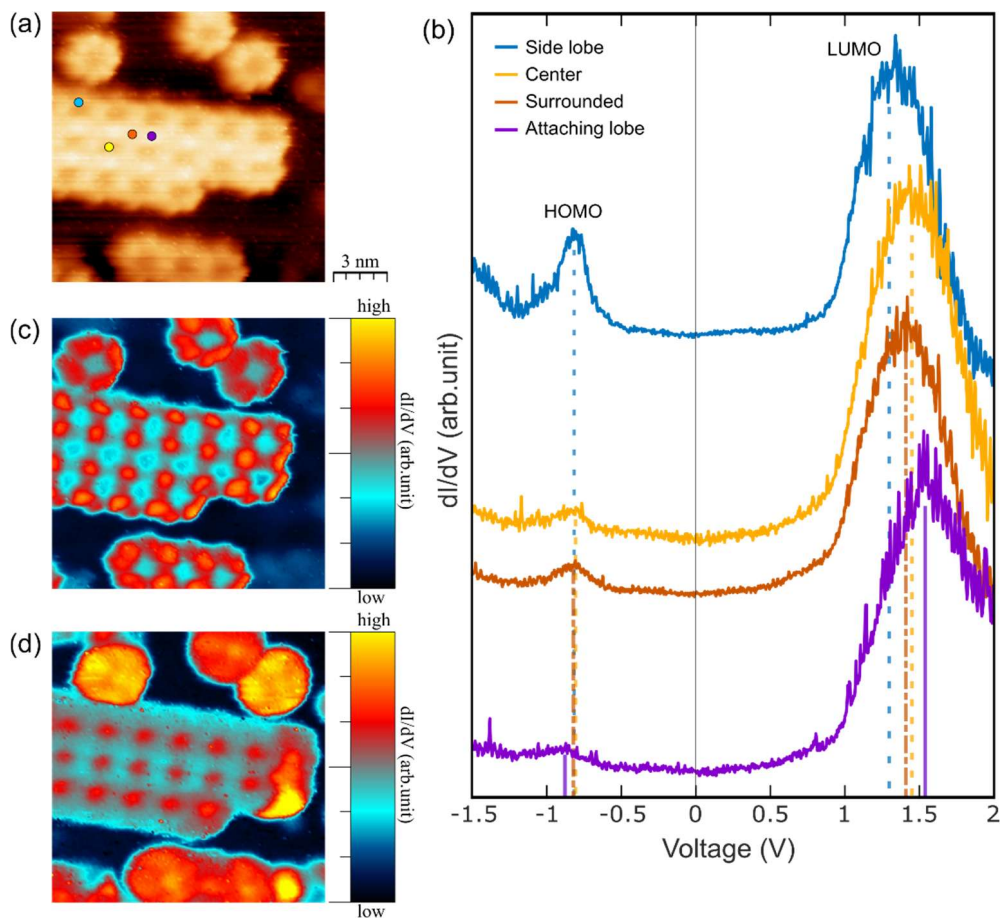

**Figure 3** Electronic properties of YPc$_2$ molecular patch on Cu(111). (a) STM image (15 × 15 nm, $V_{DC}$ = -0.7 V, $I_{set}$ = 50 pA) of self-assembled YPc$_2$ molecular patch on Cu(111). (b) dI/dV spectra with the initial condition: $V_{DC}$ = 2.0 V, $I_{set}$ = 100 pA, $V_{AC}$ = 50 mV. Spectra at the side lobe (blue), molecular center (yellow), surrounded point (orange), and attaching point (purple), were acquired at the position marked in (a) with dots of the corresponding color. (c) dI/dV map measured at HOMO energy, $V_{DC}$ = -0.7 V, $I_{set}$ = 50 pA, $V_{AC}$ = 50 mV. (d) dI/dV map measured at the LUMO energy, $V_{DC}$ = 1.3 V, $I_{set}$ = 50 mV. Each dI/dV map's intensity is rescaled independently for maximum contrast.

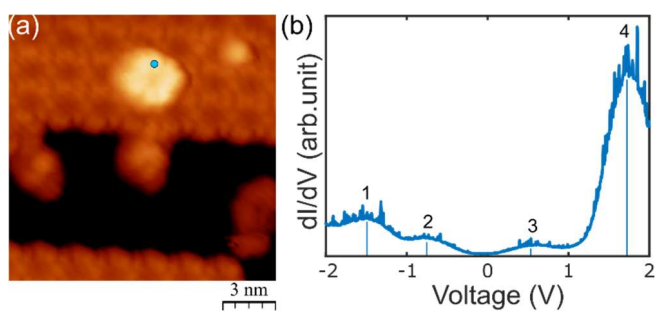

**Figure 4** Individual YPc$_2$ on YPc$_2$/Cu(111). (a) STM image (15 × 15 nm, $V_{DC}$ = 1.0 V, $I_{set}$ = 20 pA) of an individual YPc$_2$ on top of a YPc$_2$/Cu(111)$V_{DC}$ = 1.0 V. (b) dI/dV spectrum with the initial condition: $V_{DC}$ = 2.0 V, $I_{set}$ = 100 pA, $V_{AC}$ = 50 mV. Unlike YPc$_2$ adsorbed on Cu(111), we observe four peaks numbered from the lowest to the highest energy. Blue vertical lines mark the positions of the peak maxima.

## 3.2 YPc$_2$/ZnPc heterostructures on Cu(111)

The results of the previous section suggest that YPc$_2$ on Cu(111) acts as a decoupling layer for the molecules adsorbed on top of it. Nevertheless, there are some shortcomings with using YPc$_2$ as the decoupling layer. Firstly, in STM measurements, we notice that the YPc$_2$ molecules on top of YPc$_2$/Cu(111) exhibit high mobility and are easily displaced by the tip, complicating the acquisition of stable and precise data. Secondly, in view of characterizing these molecules using ensemble-averaging measurements, it is generally more convenient to introduce a decoupling layer made of a different molecular species,[16, 39] to avoid averaging over the decoupling layer containing the same metal ion but with potentially different magnetic behaviors. To circumvent these issues, we chose to use ZnPc as the diamagnetic buffer layer to decouple YPc$_2$ from the Cu(111) surface. Due to its planar structure, ZnPc is known to generate flat adsorption and self-assembly on (111) metal substrates,[40] providing a suitable template for flat adsorption of YPc$_2$. Before addressing the properties of the YPc$_2$ on ZnPc/Cu(111), we first characterize the adsorption and electronic properties of 1 ML of ZnPc on Cu(111). **Figure 5** (a) shows the self-assembled layer of ZnPc with two types of self-assembly domains labeled as A and B. Black molecular schematics of ZnPc shows the two arrangements of molecules in the layer, where the molecules in one domain are rotated by about 60° with respect to the molecules in the other domain. These different alignments reflect the symmetry of the Cu(111) substrate. A ZnPc molecule consists of the combination of two bright lobes marked with blue line and two dark lobes marked with green line forming a cross. The dI/dV spectrum of ZnPc (**Figure 5** (e)) presents three broad features marked with the vertical lines, two below and one above the Fermi level. This electronic structure is similar to ZnPc on Ag(111),[40] however, what we observe are much broader features and wider gap of 1.20 eV. This suggests a stronger hybridization is present between ZnPc and Cu(111) compared to Ag(111). The LUMO is partially occupied near Fermi level, indicating fractional charge transfer from Cu(111) and slight metal character in conductance.

To verify whether ZnPc can act as an adequate decoupling layer on Cu(111), we start investigating the structure of YPc$_2$ when deposited on top of 1 ML of ZnPc on Cu(111). To this extent, we first focus on determining the possible intermixing of the two molecules in the heterostructure, which is a common issue in ligands with very close structure. **Figure 5** (b) shows the most common type of YPc$_2$ that is observed after deposition on the single layer of ZnPc. One can see both the self-assembled monolayer of ZnPc as well as YPc$_2$ molecules. The latter can be distinguished by their apparent height, encoded as a brighter color in our STM image. The YPc$_2$ molecules that are well separated from each other show internal lobe structure, while molecules that are found in close proximity do not, indicating that the YPc$_2$-YPc$_2$ interaction affects their own electronic structure.

The apparent height of the YPc$_2$/ZnPc heterostructure, however, differs from that of the isolated YPc$_2$ on Cu(111). Comparison of the line profiles shown in **Figure 5** (c) indicates a prominence of 1.5 Å with respect to the ZnPc base layer, whose height is 2.2 Å from the Cu(111) surface. Conversely, the height of YPc$_2$ on Cu(111) is 3.7 Å, which closely corresponds to the sum of the two previous height values (**Figure 5** (c)). This result strongly points towards the possibility that YPc$_2$ molecules are physically embedded in the ZnPc layer rather than adsorbed on top of it, as sketched in **Figure 5** (d). To further verify this conjecture, we measured the dI/dV spectrum of YPc$_2$/ZnPc heterostructure (**Figure 5** (e)). Similar to YPc$_2$ on Cu(111), the dI/dV spectrum shows two prominent HOMO-LUMO peaks, with the related electronic gap depending on the molecule type, which is measured to be 2.2 eV for the YPc$_2$ molecules without inner lobes (close to one another) structure and 2.7 eV for the molecule with inner lobes structure (surrounded by ZnPc). While the former shows a HOMO-LUMO gap very similar to that of the YPc$_2$ on Cu(111), for the latter the gap is larger, possibly due to the interaction with the surrounding

ZnPc and/or modification of the dielectric environment. The similarity of the dI/dV spectra with those of YPc$_2$ on Cu(111), together with the comparison between the apparent heights in **Figure 5** (c), allows us to conclude that YPc$_2$ is embedded in the ZnPc layer. The cause of this phenomenon can be ascribed to the similarity between the YPc$_2$ and ZnPc structures, which can facilitate the intermixing between the two molecular layers at finite temperature, and to the larger adsorption energy of YPc$_2$ that favors the anchoring of this molecule to the Cu(111) surface, as it shown by DFT in **Section** 3.3.

The visibility of the inner molecular structures of YPc$_2$ is an evidence that the top Pc ligands of YPc$_2$ are sticking out from the ZnPc decoupling layer into which the YPc$_2$ molecules are embedded. When two YPc$_2$ molecules are close to each other, the overlap of the molecular lobes of the top Pc modifies the distribution of the molecular orbitals, altering the appearance in STM images. This effect is similar to that found for the phase I and phase II YPc$_2$ on Cu(111) (**Figure 1** (e) and (f)), for which the inner structure is only imaged properly when intermolecular interactions are weak (phase II).

To prevent molecular intermixing, we realized another YPc$_2$/ZnPc heterostructure with a thicker decoupling ZnPc layer. For this system, due to the absence of regions of exposed Cu(111), the thickness of the ZnPc layer could not be determined by STM images, therefore we estimated it based on the deposition time and temperature using our previous experiment as a calibration, which gives an average thickness of 2.5 ML. In this situation, we expect the coexistence of regions with 2 and 3 ML ZnPc, denoted in the following as 2-3 ML. The STM image of YPc$_2$ deposited on self-assembled 2–3 ML of ZnPc is shown in **Figure 6** (a). The thicker decoupling ZnPc layer also grows flat on Cu(111), with similar pattern as the 1 ML ZnPc case. On top of it, YPc$_2$ grows either along dislocation lines of the ZnPc layer, forming linear molecular chains, or as isolated molecules. Similar to the previous observations, the isolated molecule displays internal structure while the molecular chain does not. In addition, isolated molecules show a more pronounced two-fold symmetry, suggesting that the original four-fold symmetry is lowered due to the interaction with the underneath ZnPc (**Figure 6** (b)). On this thicker ZnPc layer, however, the apparent height of YPc$_2$ is identical to that observed for the same molecule on bare Cu(111) (**Fig. 6** (c)), suggesting the adsorption on top rather than embedded into the ZnPc layer.

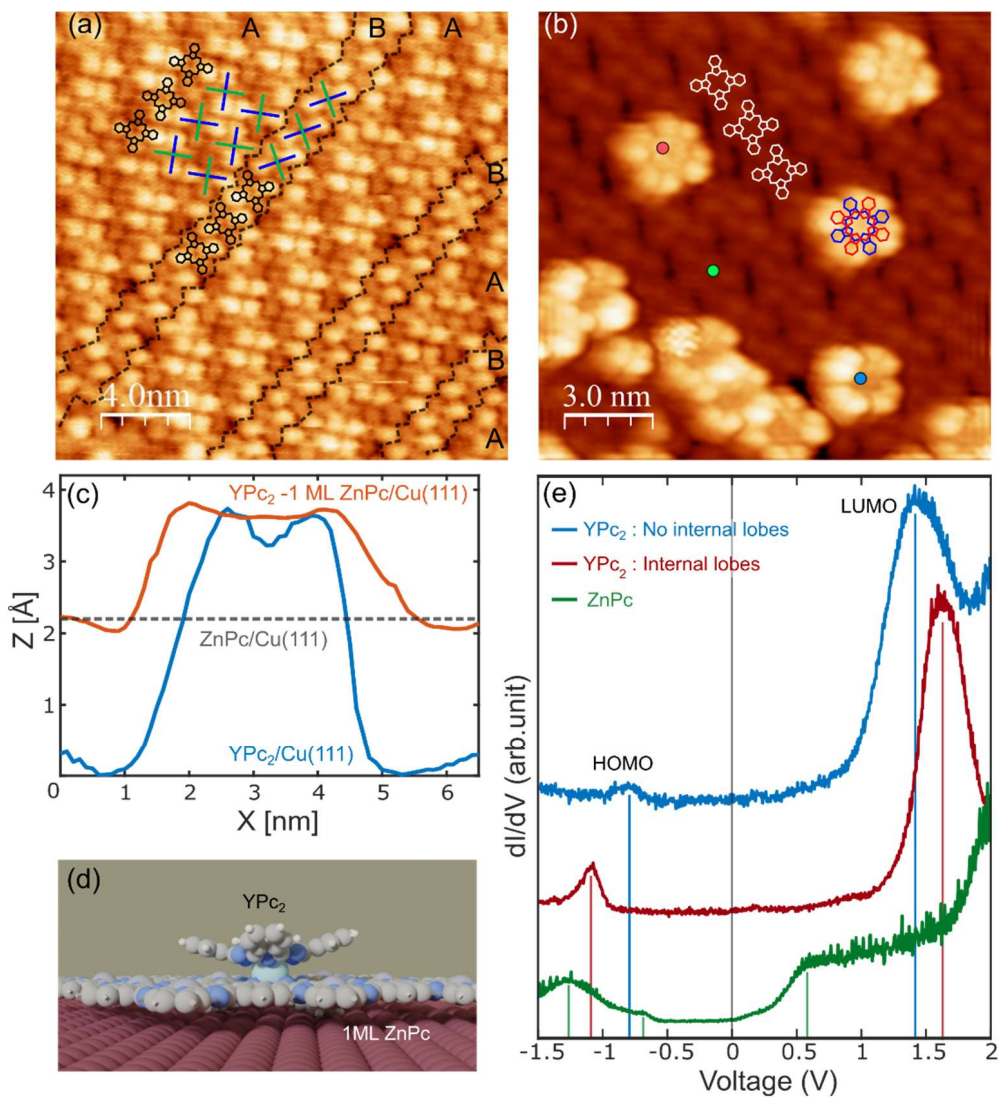

**Figure 5** Structure and electronic properties of 1ML of ZnPc and YPc$_2$-ZnPc(1 ML) heterostructures on Cu(111). (a) STM image (20 × 20 nm, V$_{DC}$ = 0.5 V, I$_{set}$ = 30 pA) of 1 ML of ZnPc. Two different orientations coexist in the same molecular patch and are marked with A and B. The corresponding schematic of the molecular structures are overlaid on the image. (b) STM image (15 × 15 nm, V$_{DC}$ = 1.5 V, I$_{set}$ = 30 pA) of YPc$_2$ deposited on the sample shown in (a). White molecular schematics indicate ZnPc, while mixed red and blue chemical structure indicates the YPc$_2$. Internal lobes were observed in spaced YPc$_2$ molecules, while more closely arranged YPc$_2$ molecules show a darker center at the same image bias. (c) Line profile of isolated YPc$_2$ (blue) and YPc$_2$ – ZnPc (1 ML) (orange). The average height of the ZnPc layer on Cu(111) is marked with a gray horizontal line. The comparison between the two line profiles suggests intermixing between the two molecular species, sketched in (d). The slight difference in lateral extension of the YPc$_2$ molecules in the two systems could be due to tip convolution effects or to a different spatial extension of the molecular orbitals. (e) dI/dV spectra comparison of ZnPc on Cu(111) and YPc$_2$ – ZnPc (1 ML) on Cu(111). Two distinct HOMO and LUMO peaks were observed for YPc$_2$ with their maxima marked with vertical solid lines, while the electronic structure of 1 ML ZnPc only shows broad features marked with solid vertical green lines.

The dI/dV spectra acquired at the molecular center and lobe (**Figure 6** (d)) show 4 peaks, similar to the YPc$_2$ on YPc$_2$/Cu(111). For both the lobe and the center, the peak maxima are positioned at about -1.5 eV (1), -0.5 eV (2),1.20 eV (2), and 2.15 eV (4). At both sites, the peak (4) is the most intense, and is reminiscent of the intense LUMO peak observed for the YPc$_2$ in contact with Cu(111). Hence, upon insertion of the ZnPc decoupling layer, the YPc$_2$/Cu(111) LUMO peak shifts approximately 0.4 eV towards higher energy Conversely, due to additional features in the dI/dV at lower energies, the electronic gap measured as (3) – (2) is 1.7 eV , which is decreased compared to the YPc$_2$ on Cu(111) and of YPc$_2$-ZnPc (1ML) on Cu(111), while closer to to that of YPc$_2$ on YPc$_2$/Cu(111).

To obtain the spatial distribution of electronic states in the YPc$_2$ adsorbed on the 2–3 ML ZnPc, we measure dI/dV map near each peak position, as shown in **Figure 6** (e)–(h). Due to the influence of the thick ZnPc decoupling layer, we observe a different pattern of dI/dV maps compared to the case of YPc$_2$ on Cu(111). The dI/dV map measured at 2.15 eV (**Figure 6** (h)) shows similarities with the LUMO distribution of YPc$_2$ on Cu(111), while **Figure 6** (e)–(g) show unclear shapes compared to dI/dV map of YPc$_2$ on Cu(111). This is presumed to be due to the interaction of these three states with the underlying ZnPc layer. In addition, despite the opposite polarity of the voltage bias, **Figure 6** (f) and (g) measured at -0.46 V and 1.0 V, respectively, show similar dI/dV distributions. The evidence that peak 2 and 3 have similar distribution of density of states suggests that they are derived from the same molecular orbital state. Summarizing, the peak 4 shows the spatial distribution and intensity of the previously labeled LUMO state, while the HOMO appears to split into two states, giving rise to the peak 2 and 3 at opposite sides of the Fermi energy.

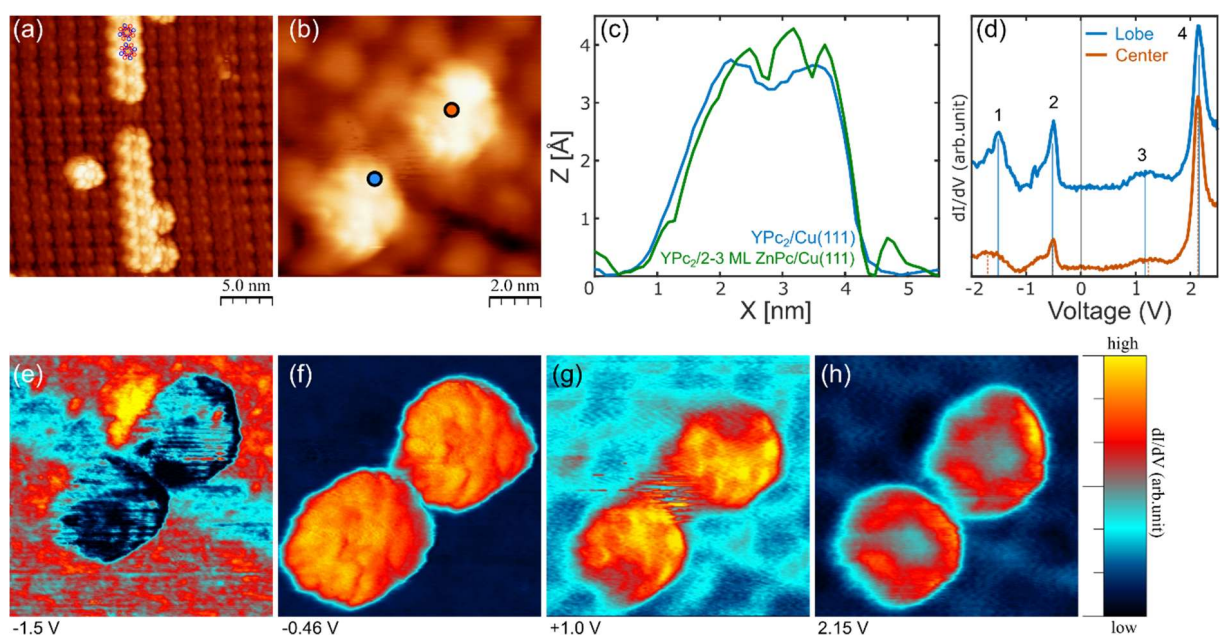

**Figure 6** Structure and electronic properties of YPc$_2$ on 2–3 ML of ZnPc on Cu(111). (a) STM image (25 nm x 25 nm, V$_{DC}$ = 1.5 V, I$_{set}$ = 30 pA) of YPc$_2$ molecules on the 2–3 ML ZnPc base layer on Cu(111). YPc$_2$ molecule adsorption configuration is marked with molecular schematics with red and blue. (b) Close-up STM image (8 × 8 nm, V$_{DC}$ = 1.0 V, I$_{set}$ = 30 pA) of two isolated YPc$_2$ molecules. (c) Line profile of YPc$_2$ on 2–3 ML of ZnPc on Cu(111) (red) compared to the isolated YPc$_2$ on Cu(111) from **Figure 5** (c) (grey). Similar height suggests that the YPc$_2$ molecule is adsorbed on top of the ZnPc layer without intermixing. (d) dI/dV spectra obtained at the molecular lobe (blue) and the molecular center (orange) at the positions marked in (b). The four peaks' maxima are marked with vertical solid lines. (e)–(h) dI/dV maps of the same area shown in (b), measured at the energy corresponding to the peaks' maxima in (c), -1.5 eV, -0.46 eV, 1.0 eV, and 2.15 eV, respectively. For all maps, I$_{set}$ = 30 pA and V$_{AC}$ = 50 mV. Each dI/dV map's intensity is rescaled independently for maximum contrast.

## 3.3 Density Functional Theory

Using DFT calculations we rationalize the change of charge and spin states of YPc$_2$ on Cu(111) and ZnPc/Cu(111). We first estimate the energy level alignment of the system by placing the respective ionization potential (IP) and electron affinity (EA) levels of ZnPc and YPc$_2$ relative to the work function of the Cu(111) substrate (see **Figure 7**). Our calculations indicate that ZnPc in vacuum is non-magnetic (**Figure 8**) and its relatively large band-gap of 2.5 eV and its IP= 5.073 eV places its electronic states well below the Fermi edge of the substrate (4.9 eV), indicating that ZnPc/Cu(111) will most likely retain the non-magnetic neutral charge state of ZnPc in vacuum. We note that the deposition of a single layer of ZnPc on Cu(111) leads to a reduction in the work function of about 0.2 eV, which is in line with the general trend observed with ZnPc on other metallic surfaces.[41] Since the ZnPc molecule remains almost perfectly planar, the reduction in work function is not a consequence of the molecular dipole. YPc$_2$ has a gap of 1.3 eV in vacuum and its IP= 4.875 eV places it closer to the Fermi edge of the substrate, indicating that YPc$_2$/Cu(111) is most likely subject to charge fluctuations and spin quenching. Our gas-phase calculations of YPc$_2$ indicate that it has an electronic configuration of Y: [Kr] 5s$^0$ 4d$^0$ Pc$_2^{1}$; *i.e.*, a non-magnetic trivalent Y$^{3+}$ and a single unpaired electron delocalized over the Pc$_2$ ligand which is split in a SOMO/SUMO state (see **Figure 8** (b)). To understand the loss of the unpaired spin when deposited directly onto Cu(111) it is instructive to compare the respective projected density of state (PDOS) plots in **Figure 8**. In vacuum, the ligand is fully polarized. When deposited on Cu(111) the ligand loses all polarization and becomes strongly broadened, indicating hybridization with the metal substrate. Furthermore, this broadening is stronger for the lower Pc ring, *i.e.*, the ring in proximity to the Cu(111) surface. When deposited on a free-standing layer of ZnPc, YPc$_2$ recovers its delocalized unpaired electron. This can be seen as the limit of a sufficiently thick ZnPc decoupling layer where YPc$_2$ is decoupled from Cu(111). The strong hybridization and resulting broadening can be an indication for the suppression of Kondo, however more detailed calculations and experiments are required to assess with certainty the presence or absence of Kondo in this system.

The intermixing of YPc$_2$ in a single layer of ZnPc seems to be driven by more than just adsorption energy differences as DFT predicts a slightly larger adsorption energy for a single YPc$_2$ molecule (-4.1 eV) when compared to a single ZnPc molecule (-3.5 eV), calculated at low coverage, and might be driven by the formation of a densely packed layer.

**Table 1** Calculated total energy [expressed in Rydeberg (Ry)] for free standing YPc$_2$, ZnPc, the related Cu(111) slabs, and the energy difference found upon adsorption of the molecules.

|  | in vacuum (Ry) | Cu 111 (Ry) | On Cu(111) (Ry) | Delta E (eV) |
|---|---|---|---|---|
| YPc$_2$ | -1974.98378748 | -36124.89789009 | -38100.18348485 | -4.1 |
| ZnPc | -1090.38354841 | -19352.64217593 | -20443.28208276 | -3.48 |

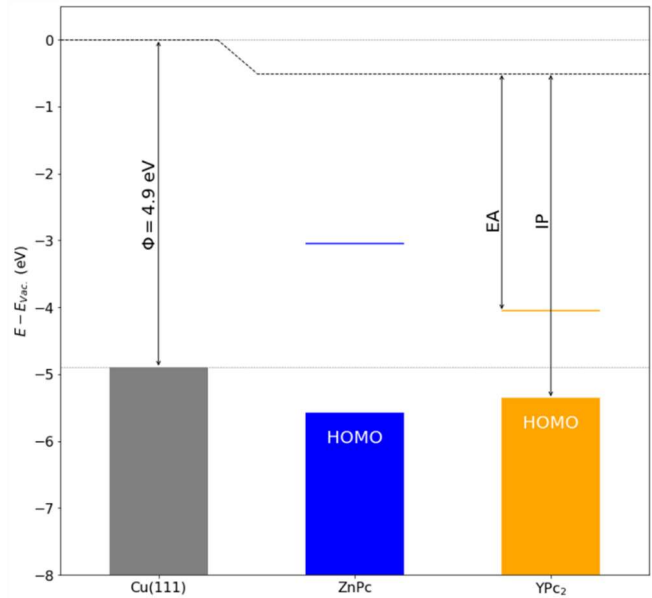

**Figure 7** Density functional theory of Cu(111), YPc$_2$, and ZnPc. Alignment of ZnPc and YPc$_2$ highest occupied levels (IP, ionization potential) and lowest unoccupied levels (EA, electron affinity) relative to the work-function of a Cu(111) surface. The vacuum level offset is due to a 0.2 eV reduction of the work-function upon deposition of the ZnPc. The wide band-gap ($E_g$=IP-EA) of ZnPc can also serve as a barrier against charge transfer from the substrate to YPc$_2$.

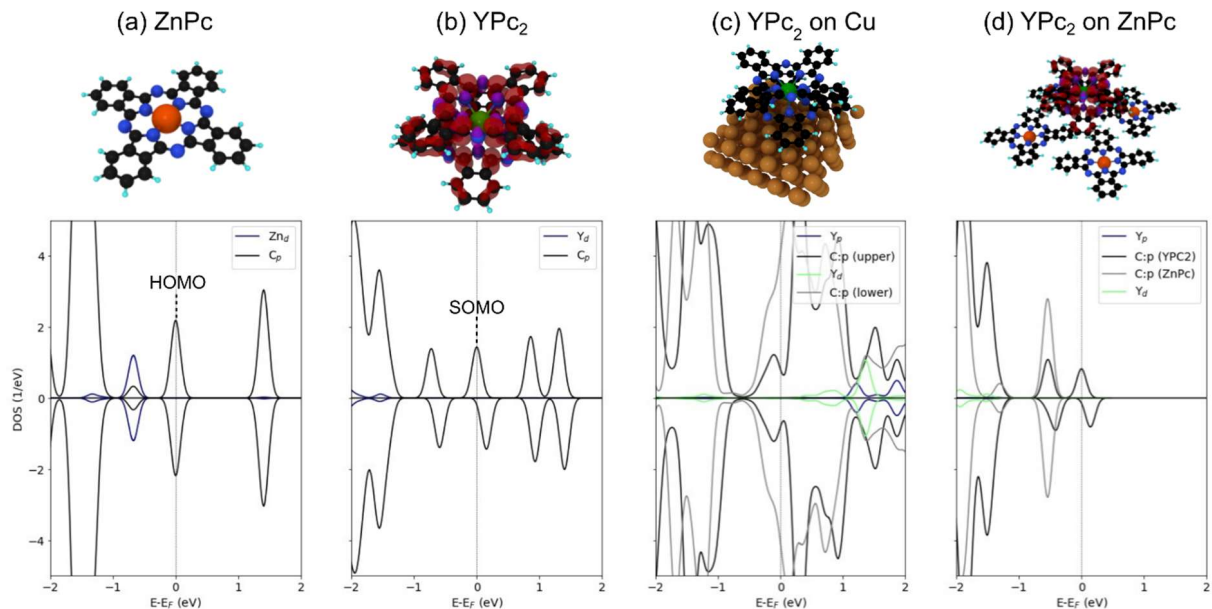

**Figure 8** Density functional theory calculations of the electronic ground-state PDOS are presented for (a) ZnPc, (b) YPc$_2$, (c) YPc$_2$ adsorbed on Cu, and (d) YPc$_2$ adsorbed on ZnPc. (a) ZnPc is non-magnetic with frontier orbitals (HOMO) of C: π type. In contrast, (b) YPc$_2$ is magnetic as shown by the red (purple) iso-surfaces (iso=±0.005 e/au$^3$) due to an unpaired electron in a C: π SOMO orbital. (c) When YPc$_2$ is adsorbed on Cu(111), the magnetic moment of YPc$_2$ is quenched leading to an S=0 molecule. The ligand carbons close to the Cu ("lower") show overall stronger hybridization however, both ligand units show non-magnetic character. (d) In such a ZnPc molecular bi-layer, no charge transfer from ZnPc to YPc$_2$ or vice versa is observed and YPc$_2$ retains its spin-polarized nature, indicating that ZnPc can be used as a buffer layer for YPc$_2$.

## 3.4 Conclusion and perspective

With this study, we identified the conditions to preserve the unpaired electron in YPc$_2$ when adsorbed on a surface and build a well-behaved molecular spin architecture. We selected ZnPc as the decoupling layer due to its single-Pc-based planar adsorption and diamagnetic properties. As only a few ML of ZnPc layer is sufficient to decouple YPc$_2$ molecules from the metal substrate, this system is suitable to be addressed with STM related magnetic characterizations such as ESR-STM.[33] In addition, the use of Cu(111) as the substrate opens up the possibility to investigate YPc$_2$ with a recently developed surface-sensitive ensemble ESR,[31] for which molecular films are typically grown on top of single crystal Cu(111)/Al$_2$O$_3$ microstrip resonators. Magnetically diluting YPc$_2$ into suitable molecular matrices may potentially allow characterizing the spin dynamics of their spins down to the limit of ultra-thin molecular films using pulsed ESR.

Finally, we identified the hybridization between YPc$_2$ and the metal substrate as the mechanism responsible for spin quenching. Further investigation of the spin properties of YPc$_2$ on other metals such as Ag and Au will allow understanding this mechanism in more depth and design robust molecular spin architectures. Finally, diamagnetic spacers made of suitable molecular structures can also enable adjusting the distance of molecular spins,[16] opening a way to the optimization not only of the molecule-substrate interaction but also of the intermolecular spin-spin coupling.

## Acknowledgements

This work was supported by the Institute for Basic Science (Grant No. IBS-R027-D1).

## References

1. A. Gaita-Arino, F. Luis, S. Hill and E. Coronado, *Nat Chem*, 2019, **11**, 301-309.
2. J. Tejada, E. M. Chudnovsky, E. d. Barco, J. M. Hernandez and T. P. Spiller, *Nanotechnology*, 2001, **12**, 181.
3. F. Troiani, V. Bellini, A. Candini, G. Lorusso and M. Affronte, *Nanotechnology*, 2010, **21**, 274009.
4. K. Bader, M. Winkler and J. van Slageren, *Chem Commun (Camb)*, 2016, **52**, 3623-3626.
5. A. Ardavan, O. Rival, J. J. Morton, S. J. Blundell, A. M. Tyryshkin, G. A. Timco and R. E. Winpenny, *Phys Rev Lett*, 2007, **98**, 057201.
6. K. Bader, D. Dengler, S. Lenz, B. Endeward, S. D. Jiang, P. Neugebauer and J. van Slageren, *Nat Commun*, 2014, **5**, 5304.
7. J. M. Zadrozny, J. Niklas, O. G. Poluektov and D. E. Freedman, *ACS Central Science*, 2015, **1**, 488-492.
8. Y. Zhang, P. Guan, H. Isshiki, M. Chen, M. Yamashita and T. Komeda, *Nano Research*, 2010,

**3**, 604-611.
9. P. Willke, T. Bilgeri, X. Zhang, Y. Wang, C. Wolf, H. Aubin, A. Heinrich and T. Choi, *ACS Nano*, 2021, **15**, 17959-17965.
10. R. Otero, J. M. Gallego, A. L. de Parga, N. Martin and R. Miranda, *Adv Mater*, 2011, **23**, 5148-5176.
11. Z. Li, Y. Li and C. Yin, *Polymers (Basel)*, 2023, **15**.
12. E. Moreno-Pineda and W. Wernsdorfer, *Photonic Quantum Technologies: Science and Applications*, 2023, **1**, 269-304.
13. L. C. de Camargo, M. Briganti, F. S. Santana, D. Stinghen, R. R. Ribeiro, G. G. Nunes, J. F. Soares, E. Salvadori, M. Chiesa and S. Benci, *Angewandte Chemie*, 2021, **133**, 2620-2625.
14. M. Briganti, G. Serrano, L. Poggini, A. L. Sorrentino, B. Cortigiani, L. C. de Camargo, J. F. Soares, A. Motta, A. Caneschi, M. Mannini, F. Totti and R. Sessoli, *Nano Lett*, 2022, **22**, 8626-8632.
15. M. Hollerer, D. Lüftner, P. Hurdax, T. Ules, S. Soubatch, F. S. Tautz, G. Koller, P. Puschnig, M. Sterrer and M. G. Ramsey, *ACS Nano*, 2017, **11**, 6252-6260.
16. K. Noh, L. Colazzo, C. Urdaniz, J. Lee, D. Krylov, P. Devi, A. Doll, A. J. Heinrich, C. Wolf, F. Donati and Y. Bae, *Nanoscale Horiz*, 2023, **8**, 624-631.
17. I. Cimatti, L. Bondì, G. Serrano, L. Malavolti, B. Cortigiani, E. Velez-Fort, D. Betto, A. Ouerghi, N. B. Brookes, S. Loth, M. Mannini, F. Totti and R. Sessoli, *Nanoscale Horizons*, 2019, **4**, 1202-1210.
18. D. Komijani, A. Ghirri, C. Bonizzoni, S. Klyatskaya, E. Moreno-Pineda, M. Ruben, A. Soncini, M. Affronte and S. Hill, *Physical Review Materials*, 2018, **2**, 024405.
19. T. Frauhammer, H. Chen, T. Balashov, G. Derenbach, S. Klyatskaya, E. Moreno-Pineda, M. Ruben and W. Wulfhekel, *Phys Rev Lett*, 2021, **127**, 123201.
20. Y. Zhang, P. Liao, J. Kan, C. Yin, N. Li, J. Liu, Q. Chen, Y. Wang, W. Chen, G. Q. Xu, J. Jiang, R. Berndt and K. Wu, *Physical Chemistry Chemical Physics*, 2015, **17**, 27019-27026.
21. E. Moreno-Pineda, C. Godfrin, F. Balestro, W. Wernsdorfer and M. Ruben, *Chem Soc Rev*, 2018, **47**, 501-513.
22. R. Barhoumi, A. Amokrane, S. Klyatskaya, M. Boero, M. Ruben and J.-P. Bucher, *Nanoscale*, 2019, **11**, 21167-21179.
23. A. Candini, D. Klar, S. Marocchi, V. Corradini, R. Biagi, V. De Renzi, U. del Pennino, F. Troiani, V. Bellini, S. Klyatskaya, M. Ruben, K. Kummer, N. B. Brookes, H. Huang, A. Soncini, H. Wende and M. Affronte, *Scientific Reports*, 2016, **6**, 21740.
24. F. Branzoli, M. Filibian, P. Carretta, S. Klyatskaya and M. Ruben, *Physical Review B*, 2009, **79**, 220404.
25. F. Branzoli, P. Carretta, M. Filibian, M. J. Graf, S. Klyatskaya, M. Ruben, F. Coneri and P. Dhakal, *Physical Review B*, 2010, **82**, 134401.
26. M. Studniarek, C. Wäckerlin, A. Singha, R. Baltic, K. Diller, F. Donati, S. Rusponi, H. Brune, Y. Lan, S. Klyatskaya, M. Ruben, A. Seitsonen and J. Dreiser, *Advanced Science*, 2019, **6**, 1901736.

27. Z. Deng, S. Rauschenbach, S. Stepanow, S. Klyatskaya, M. Ruben and K. Kern, *Physica Scripta*, 2015, **90**, 098003.
28. J. Siewert, R. Fazio, G. M. Palma and E. Sciacca, *Journal of Low Temperature Physics*, 2000, **118**, 795-804.
29. F. Branzoli, P. Carretta, M. Filibian, S. Klyatskaya and M. Ruben, *Physical Review B*, 2011, **83**, 174419.
30. A. K. Boudalis, J.-E. Olivares-Peña, E. Moreno-Pineda, A. Fediai, W. Wenzel, P. Turek and M. Ruben, *Chemical Communications*, 2021, **57**, 11505-11508.
31. F. H. Cho, J. Park, S. Oh, J. Yu, Y. Jeong, L. Colazzo, L. Spree, C. Hommel, A. Ardavan, G. Boero and F. Donati, *Review of Scientific Instruments*, 2024, **95**.
32. J. Rocker, D. Cornu, E. Kieseritzky, A. Seiler, O. Bondarchuk, W. Hansel-Ziegler, T. Risse and H. J. Freund, *Rev Sci Instrum*, 2014, **85**, 083903.
33. S. Baumann, W. Paul, T. Choi, C. P. Lutz, A. Ardavan and A. J. Heinrich, *Science*, 2015, **350**, 417-420.
34. L. Ruan, J. Tong, G. Qin, L. Zhou, X. Jiao and X. Zhang, *European Journal of Inorganic Chemistry*, 2020, **2020**, 2112-2117.
35. M. Capsoni, A. Schiffrin, K. A. Cochrane, C. G. Wang, T. Roussy, A. Q. Shaw, W. Ji and S. A. Burke, *The Journal of Physical Chemistry C*, 2017, **121**, 23574-23581.
36. D. E. Barlow and K. W. Hipps, *The Journal of Physical Chemistry B*, 2000, **104**, 5993-6000.
37. J. Hou, Y. Wang, K. Eguchi, C. Nanjo, T. Takaoka, Y. Sainoo, R. Arafune, K. Awaga and T. Komeda, *Communications Chemistry*, 2020, **3**, 36.
38. A. Kumar, K. Banerjee and P. Liljeroth, *Nanotechnology*, 2017, **28**, 082001.
39. J. H. Lee, C. Urdaniz, S. Reale, K. J. Noh, D. Krylov, A. Doll, L. Colazzo, Y. J. Bae, C. Wolf and F. Donati, *Physical Review B*, 2024, **109**, 235427.
40. A. Benhnia, S. Watanabe, R. Tuerhong, M. Nakaya, J. Onoe and J.-P. Bucher, *Nanomaterials*, 2021, **11**, 1618.
41. Y. L. Huang, E. Wruss, D. A. Egger, S. Kera, N. Ueno, W. A. Saidi, T. Bucko, A. T. S. Wee and E. Zojer, *Molecules*, 2014, **19**, 2969-2992.